\documentclass[aps,prd,nofootinbib,amsmath,amssymb,superscriptaddress,twocolumn,10pt]{revtex4}

\pdfoutput=1

\usepackage{graphicx}
\usepackage{dcolumn}
\usepackage{bm}
\usepackage{amssymb}
\usepackage{latexsym}
\usepackage{booktabs}
\usepackage{amsmath}
\usepackage{multirow}
\usepackage[colorlinks=true, linkcolor=red, citecolor=blue]{hyperref}

\newcommand{\be}{\begin{equation}}
\newcommand{\ee}{\end{equation}}
\newcommand{\bq}{\begin{eqnarray}}
\newcommand{\eq}{\end{eqnarray}}

\begin{document}

\title{Impacts of dark energy on weighing neutrinos after Planck 2015}

\author{Xin Zhang}
\email{zhangxin@mail.neu.edu.cn}
\affiliation{Department of Physics, College of Sciences,
Northeastern University, Shenyang 110004, China}
\affiliation{Center for High Energy Physics, Peking University, Beijing 100080, China}

\begin{abstract}
We investigate how dark energy properties impact the cosmological limits on the total mass of active neutrinos. We consider two typical, simple dark energy models (that have only one more additional parameter than $\Lambda$CDM), i.e., the $w$CDM model and the holographic dark energy (HDE) model, as examples, to make an analysis. In the cosmological fits, we use the Planck 2015 temperature and polarization data, in combination with other low-redshift observations, including the baryon acoustic oscillations, type Ia supernovae, and Hubble constant measurement, as well as the Planck lensing measurements. We find that, once dynamical dark energy is considered, the degeneracy between $\sum m_\nu$ and $H_0$ will be changed, i.e., in the $\Lambda$CDM model, $\sum m_\nu$ is anti-correlated with $H_0$, but in the $w$CDM and HDE models, $\sum m_\nu$ becomes positively correlated with $H_0$. Compared to $\Lambda$CDM, in the $w$CDM model the limit on $\sum m_\nu$ becomes much looser, but in the HDE model the limit becomes much tighter. In the HDE model, we obtain $\sum m_\nu<0.113$ eV (95\% CL) with the combined data sets, which is perhaps the most stringent upper limit by far on neutrino mass. Thus, our result in the HDE model is nearly ready to diagnose the neutrino mass hierarchy with the current cosmological observations.

\end{abstract}

 \maketitle

The solar and atmospheric neutrino experiments have shown that neutrinos are massive and that there is significant mixing between different neutrino species (see \cite{Lesgourgues:2006nd} for a review). However, the measurement of the absolute neutrino mass scale is a challenge for experimental particle physics. The neutrino oscillation experiments can only measure the squared mass differences between the neutrino mass eigenstates. Cosmological observations are, nevertheless, more prone to be capable of detecting the effects of the absolute neutrino mass. The cosmic microwave background (CMB) observations, combined with large-scale structure and distance measurements, have been providing tight limits on the total mass of neutrinos (see, e.g., \cite{Ade:2013zuv,Ade:2015xua}, and references therein).

Massive neutrinos could affect the CMB anisotropies and matter clustering, thus providing a potential way to weigh them through the CMB and large-scale structure observations. The fact that neutrinos have masses changes the redshift of matter-radiation equality epoch $z_{\rm eq}$, and thus affects the position and amplitude of the acoustic peaks in the CMB power spectrum. The neutrino mass will change the angular diameter distance to the last scattering surface, $D_A(z_\ast)$. To ensure the same observed acoustic peak scale $\theta_\ast$, the effect on the background cosmology can be compensated by changing other background parameters, such as the Hubble constant $H_0$ and the equation-of-state parameter of dark energy $w$ (or other parameters characterizing dark energy properties). This is why the neutrino mass $\sum m_\nu$ (summed over the three neutrino families) is degenerated with $H_0$ and $w$. The neutrino mass also affects the slope of the CMB power spectrum at low multipoles due to the integrated Sachs-Wolfe (ISW) effect that describes the energy change of CMB photons caused by the decay of the gravitational potentials during radiation domination (early ISW effect) or dark energy domination (late ISW effect). In addition, due to the large thermal velocity of neutrinos which leads to a large free-streaming scale, the massive neutrinos could free-stream out of matter perturbations, and thus introduce a scale-dependent suppression of the clustering amplitude. This also affects the late-time effect of lensing on the CMB power spectrum, i.e., massive neutrinos also suppress the lensing potential.

The Planck 2015 CMB data have provided the tight limits on the total mass of active neutrinos, $\sum m_\nu$ \cite{Ade:2015xua}. The base $\Lambda$CDM model assumes a normal mass hierarchy of neutrinos with $\sum m_\nu\approx 0.06$ eV (dominated by the heaviest neutrino mass eigenstate). When the model is extended to allow for larger neutrino masses, a reasonable assumption is that the three species of neutrinos have degenerate masses, neglecting the small differences between mass eigenstates. For the $\Lambda$CDM model, the Planck data (Planck TT+lowP) give the constraint $\sum m_\nu<0.72$ eV.\footnote{The CMB data alone can only loosely constrain the total neutrino mass. For dynamical dark energy models, the situation is almost the same. For example, for the $w$CDM model and the holographic dark energy model, our numerical calculations show that the Planck TT+lowP data give $\sum m_\nu<0.76$ eV and $\sum m_\nu<0.61$ eV, respectively.} Here, ``lowP'' denotes the Planck low-$\ell$ temperature-polarization data. It is known that the CMB data alone have a limitation to constrain the neutrino mass due to the acoustic scale degeneracy with $H_0$. So, it is necessary to combine the CMB data with other late-time cosmological probes in order to break the degeneracy. Adding the baryon acoustic oscillation (BAO) data could help to break the acoustic scale degeneracy and tighten the constraint on $\sum m_\nu$ substantially. The Planck TT+lowP+BAO data combination changes the limit to: $\sum m_\nu<0.21$ eV. Since the full Planck mission released the first analysis of the Planck polarization data, one could add the polarization data to the constraint, which will further tighten the neutrino mass limit. The combination of Planck TT, TE, EE+lowP+BAO leads to the limit $\sum m_\nu<0.17$ eV. Note that all the upper limit values for neutrino mass quoted in this paper refer to the 95\% confidence level (CL).

It should also be noticed that dark energy properties would impact the constraints on neutrino mass with cosmological observations. The cosmological limits on neutrino mass in some dynamical dark energy models have been discussed in the literature \cite{Li:2012vn,Wang:2012uf,Zhang:2014ifa,Zhang:2015rha}. In this paper, we will investigate how a dynamical dark energy impacts the measurements of neutrino mass $\sum m_\nu$ with the Planck 2015 data, compared to the case of $\Lambda$CDM.

Though the cosmological constant (or vacuum energy) $\Lambda$ is the simplest candidate of dark energy and the $\Lambda$CDM model can fit various cosmological observations quite well, it always suffers from the severe theoretical challenges, and actually some other possibilities for dark energy could not be excluded currently. But there exists too many seemingly viable dark energy models originating from various physical considerations. Of course, in practice, it is not possible, and not necessary, to discuss them one by one. A feasible scheme is to choose some typical dark energy models, as the simple extensions to $\Lambda$CDM, to detect the effects of dark energy properties on weighing neutrinos.

In this paper we only consider the simplest dynamical dark energy models, which means that the models we choose are those having only one more parameter compared to $\Lambda$CDM. The first one is the $w$CDM model in which dark energy has a constant equation of state (EoS) $w$. Though it is simple, it seems that we have no reason to let $w$ remain constant in the actual physical consideration. Usually, one can test a time-varying EoS by adopting the parametrization of $w(a)=w_0+w_a(1-a)$. But this will introduce one more additional parameter, and so we do not consider such a case in this paper. Actually, a more reasonable way in this regard to test a time-varying EoS is to consider the holographic dark energy (HDE) model \cite{Li:2004rb,Huang:2004ai,Zhang:2014ija} which has the only additional parameter $c$ in the definition of its energy density, $\rho_{\rm de}=3c^2 M_{\rm Pl}^2 R_{\rm EH}^{-2}$, where $M_{\rm Pl}$ is the reduced Planck mass and $R_{\rm EH}$ is the event horizon size of the universe. Note here that $c$ is not the speed of light (actually we adopt in this paper the natural units in which the speed of light equals one), but the parameter of HDE. In the HDE model, $c$ is a dimensionless parameter that solely determines the evolution of dark energy \cite{Zhang:2005hs,Zhang:2007sh,Zhang:2006av,Zhang:2006qu,Li:2009bn}; see the equations (2.4)--(2.7) in \cite{Zhang:2015rha} for the evolution of dark energy in the HDE model with massive neutrinos and dark radiation.

We use the Planck 2015 CMB power spectra data, in combination with other astrophysical data, to place constraints on the neutrino mass in the two considered dynamical dark energy models, and then make a comparison with the case of $\Lambda$CDM. We use the Planck 2015 CMB temperature and polarization data \cite{Aghanim:2015xee} in our calculations. We consider the combination of the likelihood at $30\leq \ell\leq 2500$ in the TT, TE, and EE power spectra and the Planck low-$\ell$ likelihood in the range of $2\leq\ell\leq 29$, which is denoted as ``Planck TT,TE,EE+lowP,'' following the nomenclature of the Planck collaboration \cite{Ade:2015xua}. In order to break the geometric degeneracy, it is necessary to consider the BAO data. Following \cite{Ade:2015xua}, we use the BAO data of the six-degree-field galaxy survey (6dFGS) at $z_{\rm eff}=0.106$ \cite{Beutler:2011hx}, the SDSS main galaxy sample (MGS) at $z_{\rm eff}=0.15$ \cite{Ross:2014qpa}, the baryon oscillation spectroscopic survey (BOSS) ``LOWZ'' at $z_{\rm eff}=0.32$ \cite{Anderson:2013zyy}, and the BOSS CMASS (i.e., ``constant mass'' sample) at $z_{\rm eff}=0.57$ \cite{Anderson:2013zyy}. Our basic data combination adopted in this paper is the Planck TT,TE,EE+lowP+BAO combination.

To simultaneously constrain dark energy parameters, one needs to include more low-redshift measurements. We thus consider the type Ia supernova (SN) data and the Hubble constant measurement. For the SN data, we use the ``joint light-curve analysis'' (JLA) sample \cite{Betoule:2014frx}. For the Hubble constant direct measurement, we use the value given by Efstathiou \cite{Efstathiou:2013via}, $H_0=70.6\pm3.3$ km s$^{-1}$ Mpc$^{-1}$ (derived from a re-analysis of the Cepheid data of Riess et al. \cite{Riess:2011yx} using the revised geometric maser distance to NGC 4258). In addition, since the CMB lensing can provide additional information at low redshifts, it is also useful to employ the Planck lensing likelihood \cite{Ade:2015zua} in our calculations.

In addition, to probe the neutrino mass and dark energy properties, the measurements of growth of structure are rather important. For example, the observations of redshift space distortions (RSD), weak gravitational lensing (WL), and galaxy cluster counts have been used to search for massive neutrinos \cite{Zhang:2014dxk,Dvorkin:2014lea,Zhang:2014nta,Li:2014dja,Archidiacono:2014apa,Bergstrom:2014fqa,Leistedt:2014sia,Beutler:2014yhv,Mantz:2014paa,DiValentino:2015wba,Rossi:2014nea,DiValentino:2015sam} and to distinguish between effects of dark energy and modified gravity \cite{Ade:2015rim,Samushia:2012iq,Beutler:2013yhm,Zhang:2014lfa,Li:2015poa}, and the combination of Planck, RSD and WL data does prefer extensions to the base $\Lambda$CDM cosmology. However, it is believed that currently significant, uncontrolled systematics still remains, more or less, in these measurements. Therefore, we do not use RSD, WL, or cluster counts measurements for combined constraints in this paper.

We use the {\tt CosmoMC} package \cite{Lewis:2002ah} to infer the posterior probability distributions of parameters. We set flat priors for the base parameters; the prior ranges for the parameters are chosen to be much wider than the posterior in order not to affect the results of parameter estimation. The perturbations in dark energy are also considered in our calculations, which is physically necessary, though for smooth dark energy the clustering of dark energy inside the horizon is strongly suppressed. To deal with the perturbation divergence problem at the $w=-1$ crossing \cite{Zhao:2005vj} in some dynamical dark energy models such as the HDE model, we employ the ``parametrized post-Friedmann'' (PPF) framework \cite{Hu:2007pj,Hu:2008zd} as implemented in {\tt CAMB} \cite{Fang:2008sn} (see also \cite{Li:2014eha,Li:2014cee,Li:2015vla}). In the following, we report the results of the parameter estimation.

\begin{table*}
\caption{Fit results for the neutrino extended models of $\Lambda$CDM, $w$CDM, and HDE cosmologies. Best fit values with $\pm 1\sigma$ errors are presented, but for the total neutrino mass $\sum m_\nu$, the 95\% upper limits are given.}
\centering
\begin{tabular}{ccccccccc}\hline
\hline Data &\multicolumn{3}{c}{Planck TT,TE,EE+lowP+BAO}&&\multicolumn{3}{c}{Planck TT,TE,EE+lowP+BAO+lensing+SN+$H_{0}$}&\\
           \cline{2-4}\cline{6-8}

  Model & $\Lambda$CDM &$ w$CDM&HDE &&$\Lambda$CDM &$w$CDM &HDE\\

\hline

$\Omega_{\rm b}h^2$&$0.02228\pm0.00015$&$0.02223^{+0.00016}_{-0.00015}$&$0.02228\pm0.00015$&&$0.02229\pm0.00014$&$0.02226\pm0.00015$&$0.02237\pm0.00015$\\
$\Omega_{\rm c}h^2$&$0.1192\pm0.0011$&$0.1197\pm0.0014$&$0.1191\pm0.0013$&&$0.1187\pm0.0011$&$0.119\pm0.0012$&$0.1177^{+0.0011}_{-0.0012}$\\
$100\theta_{\rm MC}$&$1.04083^{+0.00030}_{-0.00031}$&$1.04075^{+0.00032}_{-0.00031}$&$1.04086\pm0.00032$&&$1.0409\pm0.00029$&$1.04086\pm0.0003$&$1.04105\pm0.0003$\\
$\tau$&$0.082\pm0.017$&$0.081\pm0.018$&$0.086\pm0.017$&&$0.068^{+0.014}_{-0.016}$&$0.068^{+0.015}_{-0.016}$&$0.083\pm0.014$\\
$w/c$\tablenote{$w$ is for $w$CDM and $c$ is for HDE. }&...&$-1.068^{+0.077}_{-0.070}$&$0.533^{+0.048}_{-0.056}$&&...&$-1.043^{+0.056}_{-0.047}$&$0 .633^{+0.032}_{-0.039}$\\
$\sum m_\nu\,[\rm eV]$&$< 0.177$&$< 0.328$&$< 0.168$&&$<0.197$&$<0.304$&$<0.113$\\
$n_{\rm s}$&$0.9659\pm0.0041$&$0.9645\pm0.0046$&$0.9663\pm0.0045$&&$0.9669^{+0.0041}_{-0.0040}$&$0.9656\pm0.0043$&$0.9697\pm0.0044$\\
${\rm{ln}}(10^{10}A_{\rm s})$&$3.097\pm0.033$&$3.096^{+0.035}_{-0.034}$&$3.105^{+0.032}_{-0.033}$&&$3.067^{+0.027}_{-0.029}$&$3.067^{+0.027}_{-0.031}$&$3.096^{+0.027}_{-0.026}$\\
\hline
$\Omega_{\rm m}$&$0.3128^{+0.0073}_{-0.0075}$&$0.304\pm0.014$&$0.276^{+0.016}_{-0.015}$&&$0.3109^{+0.0070}_{-0.0079}$&$0.3068^{+0.0092}_{-0.0093}$&$0.3008^{+0.0090}_{-0.0098}$\\
$H_0$&$67.6\pm0.6$&$68.7^{+1.6}_{-1.9}$&$71.9^{+2.0}_{-2.4}$&&$67.55^{+0.64}_{-0.56}$&$68.2\pm1.0$&$68.4\pm1.0$\\
$\sigma_8$&$0.829^{+0.019}_{-0.016}$&$0.836\pm0.023$&$0.862\pm0.025$&&$0.811^{+0.015}_{-0.011}$&$0.813^{+0.017}_{-0.014}$&$0.818\pm0.013$\\
\hline
$\chi^{2}_{\rm min}$ &12940.94 &12939.28 &12945.50 && 13659.04&13655.89 &13671.50\\
\hline
\hline
\end{tabular}
\label{tab1}
\end{table*}

\begin{figure*}
\includegraphics[width=17cm]{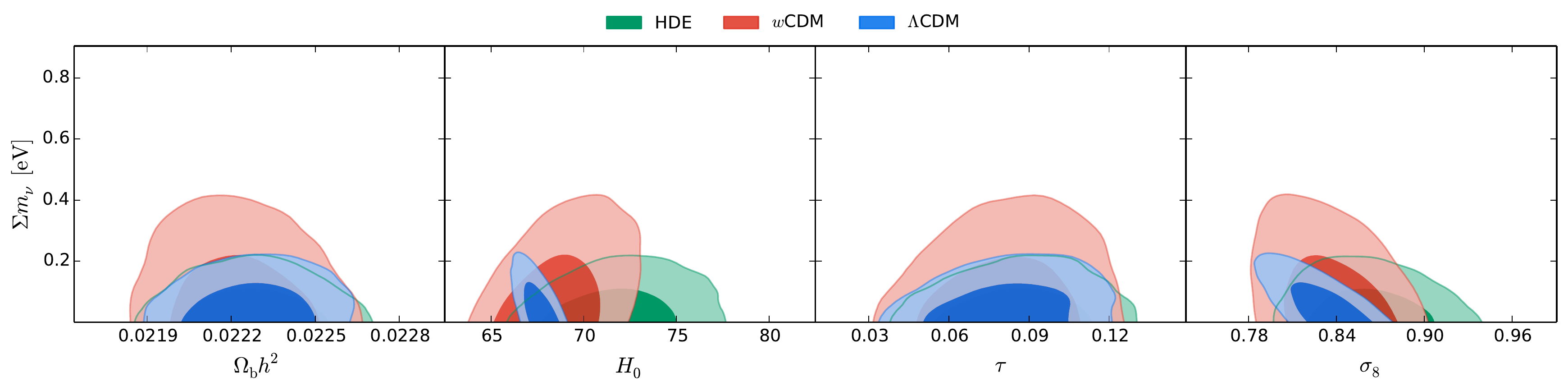}
\caption{The Planck TT,TE,EE+lowP+BAO constraints on the $\Lambda$CDM (blue), $w$CDM (red), and HDE (green) models. The 68\% and 95\% confidence level contours are shown in the parameter planes of $\sum m_\nu$ versus $\Omega_{\rm b}h^2$, $H_0$, $\tau$, and $\sigma_8$.}
\label{fig1}
\end{figure*}

\begin{figure*}
\includegraphics[width=17cm]{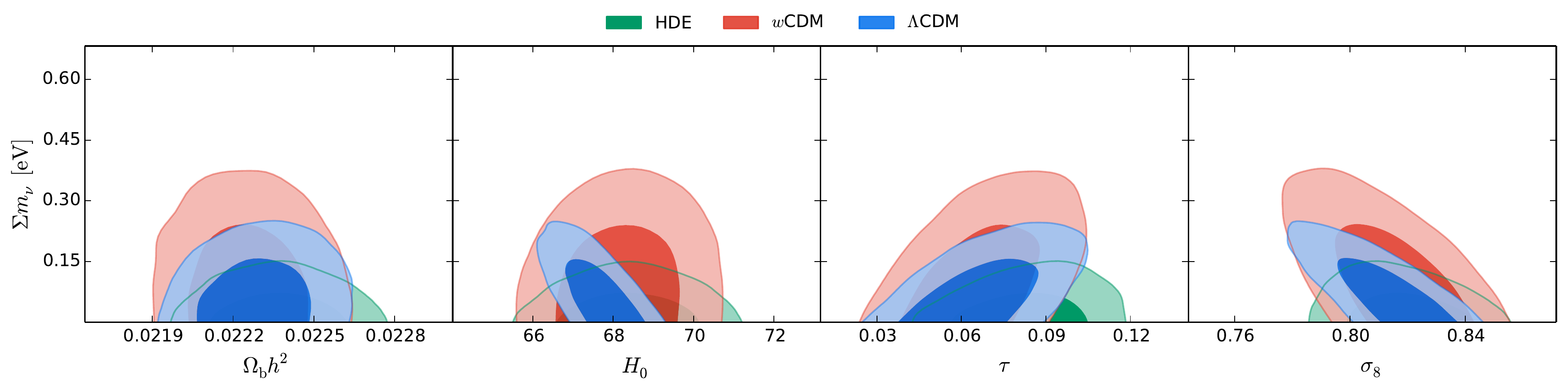}
\caption{The Planck TT,TE,EE+lowP+BAO+lensing+SN+$H_{0}$ constraints on the $\Lambda$CDM (blue), $w$CDM (red), and HDE (green) models. The 68\% and 95\% confidence level contours are shown in the parameter planes of $\sum m_\nu$ versus $\Omega_{\rm b}h^2$, $H_0$, $\tau$, and $\sigma_8$.}
\label{fig2}
\end{figure*}

\begin{figure*}
\includegraphics[width=6cm]{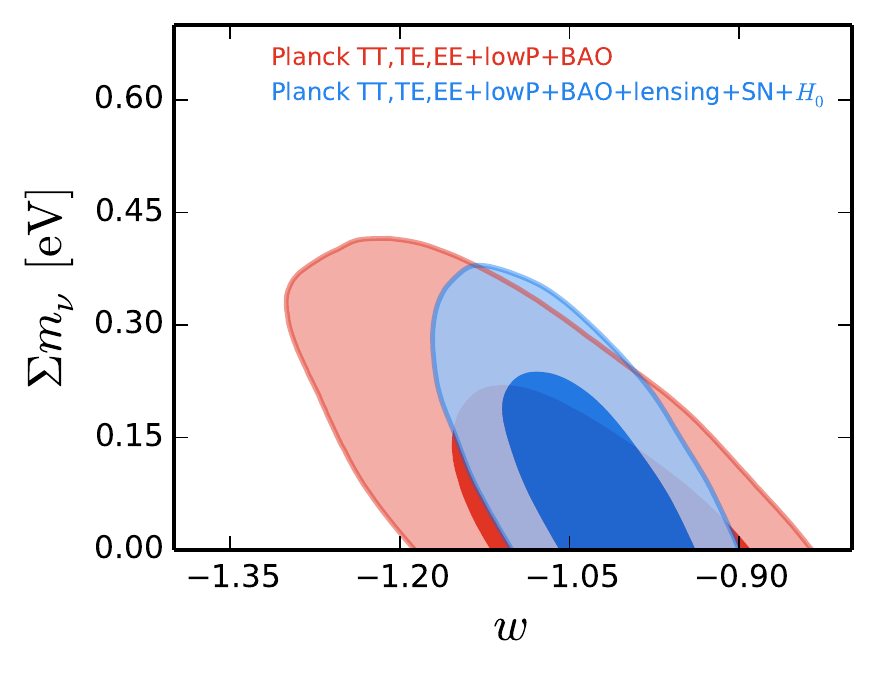}
\includegraphics[width=6cm]{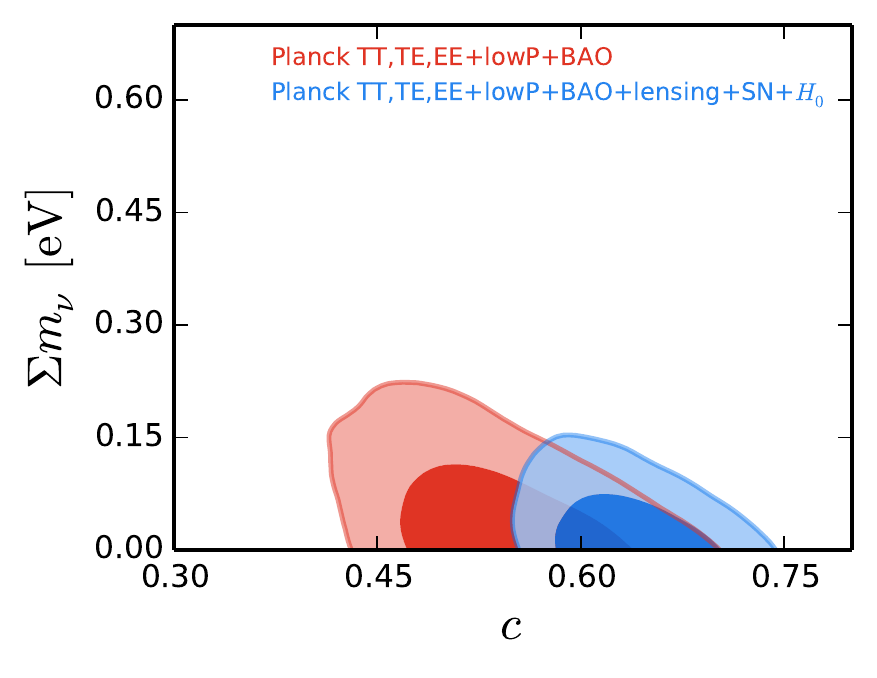}
\caption{Two-dimensional joint, marginalized constraints (68\% and 95\% confidence level) on the $w$CDM and HDE models from the Planck TT,TE,EE+lowP+BAO (red) and Planck TT,TE,EE+lowP+BAO+lensing+SN+$H_{0}$ (blue) data combinations. The constraint results in the $\sum m_\nu$--$w$ (for $w$CDM, {\it left} panel) and $\sum m_\nu$--$c$ (for HDE, {\it right} panel) planes are shown.}
\label{fig3}
\end{figure*}


We discuss the fitting results of the three models, $\Lambda$CDM, $w$CDM, and HDE, under the constraints from the two data combinations, i.e., Planck TT, TE, EE + lowP + BAO and Planck TT, TE, EE + lowP + BAO + lensing + SN + $H_0$. In these models, we consider the total mass of three species of active neutrinos, $\sum m_\nu$. The degenerate-mass assumption is made for the three species of neutrinos, as mentioned above. Note also that the three cosmological models have different numbers of parameters: the $\Lambda$CDM model has seven base parameters, and the $w$CDM and HDE models have eight base parameters. 

The fitting results are shown in Table \ref{tab1} and Figs. \ref{fig1}--\ref{fig3}. In Table \ref{tab1}, we present the fit values with $\pm1\sigma$ errors for the parameters, but for the total mass of neutrinos $\sum m_\nu$, we give the 95\% CL upper limits. In Figs. \ref{fig1}--\ref{fig3}, we show the 68\% and 95\% CL posterior distribution contours for the models.

In the case of Planck TT,TE,EE+lowP+BAO constraints, we have $\sum m_\nu<0.177$ eV for $\Lambda$CDM, $\sum m_\nu<0.328$ eV for $w$CDM, and $\sum m_\nu<0.168$ eV for HDE. We find that in the $w$CDM model, the upper limit of $\sum m_\nu$ is much bigger than that in the $\Lambda$CDM model. But in this case, the HDE model yields a slightly smaller limit value of neutrino mass than the $\Lambda$CDM model. Comparing the $\chi^2$ values of the three models in the fit, we find that the $w$CDM model performs slightly better than the $\Lambda$CDM model by $\Delta\chi^2=-1.66$, at the expense of adding one more parameter. But the HDE model performs worse than $\Lambda$CDM by $\Delta\chi^2=4.56$ though it has one more parameter. A careful check reveals that the reason for this is that the HDE model cannot fit the BAO point at $z_{\rm eff}=0.57$ well in the global fit, of which $\Delta\chi^2$ contributes solely about 3.

In Fig.~\ref{fig1}, we compare the constraint results of the three models in the fit to the Planck TT,TE,EE+lowP+BAO data combination. We find that the fit values of $\Omega_{\rm b}h^2$ and $\Omega_{\rm c}h^2$ are similar for the three models, but the results of $\Omega_{\rm m}$, $H_0$, and $\sigma_8$ are quite different. This indicates that the dark energy properties play an important role in changing the fit results. 

The Planck data have accurately measured the acoustic peaks and thus the observed angular size of acoustic scale $\theta_\ast=r_s/D_A$ is determined to a high precision (much better than 0.1\% precision at 1$\sigma$). This places tight constraints on some combinations of the cosmological parameters that determine $r_s$ and $D_A$. In the cosmological fit using the Planck data, the parameter combinations must be constrained to be close to a surface of constant $\theta_\ast$. This surface depends on the models that are assumed. In the $\Lambda$CDM model, the precise determination of $\theta_\ast$ yields a nearly constant $\Omega_{\rm m}h^3$. But for the $w$CDM and HDE models, this is not true; the distributions of $\Omega_{\rm m}h^3$ in these two models are much broader than in the $\Lambda$CDM model \cite{Li:2013dha}. However, the three models have the similar distributions of $\Omega_{\rm m}h^2$ \cite{Li:2013dha}. The constraints on $\Omega_{\rm m}h^3$ and $\Omega_{\rm m}h^2$ lead to the fact that $H_0$ can be tightly constrained in the $\Lambda$CDM model, but cannot be so well constrained in dynamical dark energy models. Therefore, we find that dark energy properties significantly affect the determination of $\Omega_{\rm m}h^3$ and thus the value of $H_0$. 

From Fig. \ref{fig1}, we also find that $\sum m_\nu$ is anti-correlated with $H_0$ in the $\Lambda$CDM model, but is positively correlated with $H_0$ in the $w$CDM and HDE models. The degeneracy in the parameter space gives rise to consistent changes in parameters including $H_0$, $\Omega_{\rm m}$, $\sum m_\nu$, and $w$ (or $c$), so that the ratio of the sound horizon and angular diameter distance remains nearly constant. Changes in the dark energy density due to its dynamical properties (characterized by $w$ or $c$ in our cases) could have some effects on the parameters such as $H_0$ and $\sum m_\nu$ because they would change to compensate. Hence, the impacts of dark energy lead to the changes of correlation between $\sum m_\nu$ and $H_0$. 

Neutrino masses could suppress the powers of CMB, which can be compensated by increasing the amplitude of primordial spectrum $A_s$. The measurements of the CMB temperature power spectrum provide a highly accurate measurement of the combination $A_s e^{-2\tau}$, where $\tau$ is the reionization optical depth parameter, which leads to a strong degeneracy (positive correlation) between $A_s$ and $\tau$. Thus we can infer that $\sum m_\nu$ is positively correlated with $\tau$, which is confirmed in Fig. \ref{fig1} for all the three models. We find that the current CMB+BAO data prefer a high value of $\tau$ ($\tau\sim 0.081-0.086$) for all the models. As shown in \cite{Ade:2015xua}, the CMB lensing data could break the degeneracy between $A_s$ and $\tau$, and in the case of $\Lambda$CDM, the value of $\tau$ could be lowered by adding lensing data.

In addition, due to the free-streaming of massive neutrinos, larger masses tend to prefer lower $\sigma_8$, thus $\sum m_\nu$ is anti-correlated with $\sigma_8$, as confirmed by the constraint results for all the models in Fig. \ref{fig1}.

Adding the BAO data helps partly break the geometric degeneracies, but not enough. In order to further break the degeneracies, in particular, to accurately probe the dark energy properties and measure neutrino mass, one needs to add more low-redshift observations, such as SN and $H_0$, as well as CMB lensing. Figure \ref{fig2} shows the comparison of the three models under the constraints from Planck TT,TE,EE+lowP+BAO+lensing+SN+$H_0$. In this case, we have $\sum m_\nu<0.197$ eV for $\Lambda$CDM, $\sum m_\nu<0.304$ eV for $w$CDM, and $\sum m_\nu<0.113$ eV for HDE. 
Thus, we see that adding the low-redshift data changes the values of upper limit of $\sum m_\nu$. For the $\Lambda$CDM model, the constraint becomes slightly weaker; for the $w$CDM model, the constraint becomes slightly tighter; and for the HDE model, the constraint is significantly tightened.

We also find that adding the CMB lensing data indeed helps improve the measurement of $\tau$. For the $\Lambda$CDM and $w$CDM models, the values of $\tau$ are significantly lowered by adding lensing data, but for the HDE data, the central value of $\tau$ is still at 0.083 in this case.

To directly show how dark energy properties affect the constraints on neutrino mass, we plot the two-dimensional posterior distribution contours (68\% and 95\% CL) in the $\sum m_\nu$--$w$ plane for $w$CDM and in the $\sum m_\nu$--$c$ plane for HDE, under the constraints from the both two data combinations, in Fig. \ref{fig3}. We find that, in the $w$CDM model, $\sum m_\nu$ is anti-correlated with $w$, and in the HDE model, $\sum m_\nu$ is also anti-correlated with $c$. Evidently, adding low-redshift data tightens the constraints on dark energy parameters and thus changes the limits on neutrino mass. For the $w$CDM model, we have $w=-1.068^{+0.077}_{-0.070}$ from CMB+BAO, and it changes to $w=-1.043^{+0.056}_{-0.047}$ when the low-redshift data are added. For the HDE model, we have $c=0.533^{+0.048}_{-0.056}$ from CMB+BAO, and it changes to $c=0.633^{+0.032}_{-0.039}$ when the low-redshift observations are included. We find that, for the HDE model, the Planck CMB data prefer a low value of $c$, and the low-redshift observations tend to drive $c$ upward to a higher value, which is in accordance with the conclusions in previous studies \cite{Zhang:2015rha,Li:2013dha}. The changes of $c$ strikingly impact the measurements of neutrino mass, and thus we obtain the extremely stringent limit, $\sum m_\nu<0.113$ eV, in this case. Since $\sum m_\nu$ must be greater than approximately 0.1 eV in the inverted mass hierarchy (degenerate hierarchy in cosmology) \cite{GonzalezGarcia:2012sz}, our result in the HDE model is nearly ready to diagnose the neutrino mass hierarchy with the current cosmological probes.

In conclusion, the dark energy properties could significantly impact the constraint limits on neutrino mass $\sum m_\nu$. To test how dynamical dark energy would affect the constraints on $\sum m_\nu$, we employ two typical, simple dark energy models as examples, i.e., the $w$CDM model and the HDE model, to make an analysis. They both have only one more parameter than the $\Lambda$CDM model. We use the Planck 2015 temperature and polarization data, in combination with other low-redshift observations, to constrain the models. The acoustic scale degeneracy leads to consistent changes in parameters such as $H_0$, $\sum m_\nu$, and $w$ (or $c$), which ensures that the ratio of the sound horizon and angular diameter distance remains nearly constant. Thus the dark energy parameters could have effects on the parameters such as $H_0$ and $\sum m_\nu$ since they would change to compensate. This leads to the fact that, in the $\Lambda$CDM model, $\sum m_\nu$ is anti-correlated with $H_0$, but once dynamical dark energy is introduced, $\sum m_\nu$ becomes positively correlated with $H_0$, as shown in the $w$CDM and HDE models (Figs. \ref{fig1} and \ref{fig2}). Our analysis also shows that $\sum m_\nu$ is anti-correlated with $w$ in the $w$CDM model and is anti-correlated with $c$ in the HDE model (Fig. \ref{fig3}). We find that in the $w$CDM model the limits on $\sum m_\nu$ become much looser, but in the HDE model the limits become much more stringent.

Using the Planck TT,TE,EE+lowP+BAO data, we obtain $\sum m_\nu<0.177$ eV for $\Lambda$CDM, $\sum m_\nu<0.328$ eV for $w$CDM, and $\sum m_\nu<0.168$ eV for HDE. Using the Planck TT,TE,EE+lowP+BAO+lensing+SN+$H_0$ data, we obtain $\sum m_\nu<0.197$ eV for $\Lambda$CDM, $\sum m_\nu<0.304$ eV for $w$CDM, and $\sum m_\nu<0.113$ eV for HDE. Therefore, we conclude that in the HDE model we can get perhaps the most stringent upper limit by far on the total mass of active neutrinos, $\sum m_\nu<0.113$ eV, with the combined cosmological data sets. Our study shows that, if dark energy is not a cosmological constant, then the allowed neutrino mass window could become much tighter and would be further closed by forthcoming observations. We will leave further relevant discussions on this subject to a forthcoming paper \cite{zhaomm}.

\begin{acknowledgments}
The author would like to thank Yun-He Li, Jing-Fei Zhang, Ming-Ming Zhao, and Shun Zhou for helpful discussions.
This work is supported by the Top-Notch Young Talents Program of China, the National Natural Science Foundation of China (Grants No.~11522540 and No.~11175042), and the Fundamental Research Funds for the Central Universities (Grants No. N140505002, No. N140506002, and No. L1505007).

\end{acknowledgments}

\end{document}